\documentclass[acmsmall]{acmart}
\usepackage[utf8]{inputenc}
\usepackage[inline]{enumitem}
\usepackage{amsmath}
\usepackage{url}

\title{The Limits of Differential Privacy (and its Misuse in Data Release and Machine Learning)}
\author{Josep Domingo-Ferrer}
\email{josep.domingo@urv.cat}
\author{David S\'anchez}
\email{david.sanchez@urv.cat}
\author{Alberto Blanco-Justicia}
\email{alberto.blanco@urv.cat}

\affiliation{%
  	\institution{Universitat Rovira i Virgili}
        \department{Dept. of Computer Science and Mathematics}
        \department{UNESCO Chair in Data Privacy}
        \department{CYBERCAT-Center for Cybersecurity Research of Catalonia}
        \streetaddress{Av. Pa\"isos Catalans 26, 43007}
        \city{Tarragona}
        \country{Catalonia}}


\received{April 2020}

\begin{document}
\maketitle
\renewcommand{\shortauthors}{Domingo-Ferrer et al.}

The traditional approach to statistical disclosure control (SDC) for privacy 
protection is utility-first. Since the 1970s, national statistical institutes 
have been using anonymization methods with heuristic parameter
choice and suitable utility preservation properties to protect
data before release.
%~\cite{hundepool}. 
Their goal is to publish analytically useful data
that cannot be linked to specific respondents or leak confidential information 
on them. 

In the late 1990s, the computer science community took another angle
and proposed privacy-first data protection. In this approach 
a {\em privacy model} specifying an {\em ex ante} privacy condition
is enforced using one or several SDC methods, such as noise addition, 
generalization or microaggregation. 
The parameters of the SDC methods depend on the privacy model 
parameters, and too strict a choice of the latter may result
in poor utility. 
%David: una mica reescrit
The first widely-accepted privacy model
was $k$-anonymity%~\cite{samarati}
, whereas differential 
privacy (DP) is 
the model that currently attracts the most attention.

DP was originally proposed for interactive statistical queries to a database~\cite{book}. 
A randomized
query function $\kappa$ (that returns the query answer plus some 
noise) satisfies $\epsilon$-DP if for all data sets $D_1$ and $D_2$
that differ in one record 
%(neighbor data sets) 
and all 
$S \subset Range(\kappa)$, it holds that
$\Pr(\kappa(D_1) \in S) \leq \exp(\epsilon) \times \Pr(\kappa(D_2) \in S)$.
In other words, the presence or absence of any single record
must not be noticeable from the query answers, up to an exponential
factor of $\epsilon$. 
The smaller $\epsilon$, the higher the protection.
%In case each record corresponds to a different respondent, 
%no released randomized
%query answer will leak the respondent's record, which 
%ensures privacy to any individual respondent.
%David: afegit, per lligar-ho amb el problema de fer servir DP per data releases (identity queries)
The noise to be added to the answer to enforce a 
certain $\epsilon$ depends on the 
global sensitivity of the query to the presence or absence of any single record.
Mild noise may suffice for statistical queries such as the mean, while 
 a very large noise is needed for identity queries returning the contents of
a specific record.

DP offers a very neat privacy guarantee and, unlike 
privacy models in the $k$-anonymity family, does not make assumptions 
on the intruder's background knowledge 
(although it assumes that all records in the database
are independent~\cite{clifton,kifer}). For this reason, 
%David: desenvolupat
DP was rapidly adopted by the research community to the point  
that previous approaches tend to be regarded as obsolete.
Researchers and practitioners have extended
the use of DP beyond the interactive setting it was designed for. 
Extended uses include: 
%($k$-anonymity-like) 
data release, where 
 privacy of respondents versus data analysts is the goal, 
%which allow more flexible analyses than 
%interactive queries, 
and collection of personal information, where privacy
of respondents versus the data collector is claimed.
Google, Apple and Facebook have seen the chance to collect 
or release microdata (individual respondent
data) from their users under the privacy pledge ``don't worry, 
whatever you tell us will be DP-protected''.

%David: afegit
However, applying DP to record-level data release 
or collection (which is equivalent to answering identity queries) 
requires employing a large amount of noise to enforce a safe enough $\epsilon$.
As a result, if $\epsilon \leq 1$ is used, as recommended in~\cite{book} to obtain a meaningful
privacy guarantee, the analytical utility of DP outputs 
is likely to be very poor~\cite{fredrikson,bambauer}.
This problem arose as soon as DP was moved outside the interactive setting.
A straightforward way to mitigate the utility problem is to use 
unreasonably large $\epsilon$.

%David: afegits detalls dels SOs i versions
Let us look at data collection.
Apple reportedly uses $\epsilon=6$ in MacOS 
and $\epsilon=14$ in iOS 10 (with some beta versions using even $\epsilon=43$)~\cite{wired}. In their RAPPOR technology, Google uses $\epsilon$ up to 9.
%at the cost of narrowing their target data uses. 
According to Frank McSherry, one of the co-inventors of DP,
using $\epsilon$ values as high as 14 is pointless in terms of 
privacy~\cite{wired}. 
%David: afegit
Indeed, the privacy guarantee of DP completely fades away 
for such large values of  $\epsilon$~\cite{book}.

As to data release, Facebook has recently 
released DP-protected data sets for social
science research, 
but it is unclear which $\epsilon$ value they 
 have used. As pointed out in~\cite{sciencemag}, this 
makes it difficult both to
%David: afegit
understand the privacy guarantees being offered 
and to assess the trustworthiness of the results obtained on the data. 
%David: reescrit una mica en un sentit més negatiu
The first released version of this DP data set had all 
demographic information about respondents
and most of the time and location information removed, and event counts 
had been added noise with $\sigma=200$~\cite{francis}. 
The second version looks better, but is still analytically poorer
than the initial version released in 2018 under utility-first
anonymization based on 
%clear data transformations: 
removal of identifiers and data aggregation.
In fact, Facebook researchers have acknowledged the difficulties 
of implementing (and verifying) DP
in real-world applications~\cite{facebook}.
%Thus, Facebook's current DP-protected data set 
%is still of dubious analytical utility.

%David: especificat
%Canviat.
The U.S. Census Bureau has also announced the use of DP to disseminate
Census 2020 results~\cite{garfinkel}. A forecast of the negative impact
of DP on the utility of the current Census is given in~\cite{santos}.
%To facilitate assessing privacy and trustworthiness,
%they ought to clearly indicate the values of $\epsilon$ being used
Additionally, Ruggles {\em et al.}~\cite{ruggles} have remarked 
that DP ``is a radical departure from established Census
Bureau confidentiality laws and precedents'': the Census
must take care of preventing respondent re-identification, 
but masking respondent characteristics 
---as DP does--- is not required. 
A more fundamental objection in~\cite{ruggles} is against the 
very idea of DP-protected microdata. 
%David: especificat
As introduced above, publishing useful record-level microdata
under DP is exceedingly difficult. This is only logical: 
releasing DP-microdata, that is, individual-level
data derived from real people, contradicts the core 
idea of DP, namely that the presence or absence of any 
individual should not be noticeable from the DP output. 

A further shortcoming arises when trying to use DP to protect continuous
data collection like Apple and Google do.
DP is subject to sequential composition: 
if a data set collected at time $t_1$ 
is DP-protected with $\epsilon_1$ and a data set collected at 
time $t_2$ on 
a non-disjoint group of respondents is DP-protected with $\epsilon_2$,
the data set obtained by composing 
the two collected data sets 
is DP-protected only with $\epsilon_1 + \epsilon_2$.  
%David: afegit
Therefore, to enforce a certain $\epsilon$ after $n$ data collections 
on the same set of
individuals, each collection should be DP-protected with $\epsilon/n$, 
thereby very substantially reducing the utility of the collected data.
Strictly speaking, it is {\em impossible} to collect DP-protected
data from a community of respondents an indefinite
number of times with a meaningful privacy guarantee. 
%David: afegit
As a remedy, Apple made the simplification 
that sequential composition only applies to the data collected on 
an individual during 
the same day but not in different days~\cite{wired}. Google 
took a different way out: they use sequential composition only for values 
that have not changed from the previous collection~\cite{accessnow}. 
Both fixes are severely flawed: the data 
of an individual collected across consecutive days or
%JOSEP. Canviat.
that may have changed \emph{still refer to the same individual}, rather
than to disjoint individuals. Ignoring this and conducting 
a systematic data collection for long periods increases the 
effective $\epsilon$ and thus reduces the effective 
level of protection by several orders of magnitude.
This issue is significantly more privacy-harming than the 
above-mentioned large values of $\epsilon$ declared by Apple and Google.

Machine learning (ML) has also seen applications of DP.
In~\cite{abadi}, DP is used to ensure that deep learning models 
do not expose private information contained in the data sets
they have been trained on. Such a privacy guarantee 
is interesting to facilitate crowdsourcing of representative
training data from individual respondents. The paper describes
the impact of sequential composition over ML training epochs 
%David: especificat
(which can be viewed as continuous data collection) on 
the effective $\epsilon$: after 350 epochs, a very large $\epsilon=20$ is attained.
To obtain usable results without rising to such a large $\epsilon$, 
the authors use the $(\epsilon,\delta)$ relaxation of DP 
%and a finer-grained composition 
that keeps $\epsilon$ 
between 2 and 8. 
%David: afegit comentari i referencies
Employing relaxations of DP to avoid the ``bad press'' of large $\epsilon$
while keeping data usable is a common workaround in recent 
literature~\cite{Triastcyn,abadi}. 
However, DP relaxations are not a free lunch: relaxed DP is not DP anymore. For example,
%Alberto: afegides cometes perquè es cita literal del llibre
with $(\epsilon,\delta)$-DP, 
%JOSEP. Arreglada notació.
``$\delta$ values on the order of $1/|D|$ (where
$|D|$ is the size of the data set) are very dangerous: they permit
preserving privacy by publishing the complete records of a small number
of database participants''~\cite{book}.
The value of $\delta$ employed in~\cite{abadi,Triastcyn} is in fact 
on the order of
$1/|D|$, thereby incurring severe risk of disclosure.
Nonetheless, despite the use of relaxations and large $\epsilon$ and $\delta$, 
the impact of DP on data utility remains significant: 
%JOSEP. Arreglat.
in~\cite{abadi} the deep learning algorithm without privacy 
protection achieves 86\% accuracy 
on the CIFAR-10 data set, but
it falls down to 73\% for $\epsilon=8$ and to 67\%
for $\epsilon=2$. 

Very recently, DP has also been proposed for 
a decentralized form of ML called \emph{federated learning}.
Federated learning allows a model manager to learn
a ML model based on data that are privately stored by 
a set of clients: in each epoch, the model manager
sends the current model to the clients, who return
to the manager a model update based on their respective 
private data sets. This 
does not require the clients to surrender
their private data to the model manager and
saves computation to the latter. However, 
model updates might leak information on the clients' private
data unless properly protected. To prevent such a leakage,
DP is applied to model updates~\cite{Wei,Triastcyn}.
%\cite{shokri} propose to protect model updates using DP.
%David: afegit
However, in addition to distorting model updates, 
using DP raises the following issues:
\begin{itemize}
\item Since model updates are protected in each epoch,
and in successive epochs they are computed on
the same (or, at least, not completely disjoint) client data,
sequential composition applies. 
This means that the  
effective $\epsilon$ grows with the number of epochs,
%David: afegit
and the effective protection decreases exponentially.
%David: afegida referencia
Therefore, reasonably useful models can only be obtained 
for meaninglessly large $\epsilon$ (such as 50-100~\cite{Wei}).
\item In the original definition of DP, a data set 
where each record contains the answer of a {\em different}
respondent is assumed. 
Then DP ensures that 
the record contributed by any single respondent
is unnoticeable from the released DP-protected output. 
This protects the privacy of any single respondent. However, when 
DP is used to protect the model update submitted by a client, all
records in the client's data set belong to the client. 
Making any single record unnoticeable is not sufficient to 
protect the client's privacy when {\em all} records 
%This would only 
%make sense if the client's private data consist of 
%answers collected from different respondents, and in that 
%case, the privacy subjects would be those individual respondents
%rather than the client. 
in the client's private data set 
are about the client, as it happens {\em e.g.}
if the client's private data contain her health-related
or fitness measurements. Thus, the DP guarantee loses
its significance in this case. 
\end{itemize}

In conclusion, DP is a neat privacy definition 
that can co-exist with certain well-defined data uses 
in the context of interactive queries. However, DP is neither
a silver bullet for all privacy problems 
%David: afegit
nor a replacement for all previous privacy models~\cite{clifton}. In fact,
extreme care should be exercised when trying to 
extend its use beyond the setting
it was designed for. 
%David: reescrit una mica
%JOSEP. Reescrit.
As we have highlighted, 
fundamental misunderstandings and blatantly flawed 
implementations pervade the application
of DP to data releases, data collection and machine learning.
These misconceptions have serious 
consequences in terms of poor privacy or poor utility
and they are driven by the insistence to 
twist DP in ways that contradict its own core idea: to
make the data of any single individual unnoticeable.

\section*{Acknowledgment and disclaimer}

We acknowledge support from: European Commission
(project H2020-871042 ``SoBigData++''),
Government of Catalonia (ICREA Acad\`emia Prize to 
the first author and grant 2017 SGR 705)
and Spanish Government (projects RTI2018-095094-B-C21
 and TIN2016-80250-R).
The authors are with the UNESCO Chair in
Data Privacy, but their views here are not
necessarily shared by UNESCO.

\bibliographystyle{acm}
 
\end{document}